\documentclass[5p]{elsarticle}

\usepackage{lineno,hyperref}
\usepackage{amsmath}
\usepackage[labelfont={bf}]{caption}
\usepackage{subcaption}
\usepackage{gensymb}
\usepackage{graphicx}
\usepackage{xcolor}
\modulolinenumbers[1]
\graphicspath{{figures/}}
\def\he{$^{3}$He}
\def\li{$^{6}$Li}
\def\gr{$\gamma$-ray}
\def\grs{$\gamma$-rays}

\begin{document}

\begin{frontmatter}

\title{Measurement of the absolute neutron beam polarization from a supermirror polarizer and the absolute efficiency of a neutron spin rotator for the NPDGamma experiment using a polarized \he\ neutron spin-filter\tnoteref{dio}}
\tnotetext[dio]{https://doi.org/10.1016/j.nima.2018.03.055}

\author[utk,mit]{M.M.~Musgrave\corref{cor1}}
\cortext[cor1]{Corresponding author}
\ead{musgrave@mit.edu,mmusgrav@vols.utk.edu}

\author[uva,ornl]{S.~Bae{\ss}ler}
\author[asu]{S.~Balascuta}
\author[udam]{L.~Barr\'{o}n-Palos}
\author[asu]{D.~Blyth}
\author[ornl]{J.D.~Bowman}
\author[umi]{T.E.~Chupp}
\author[ornl]{V.~Cianciolo}
\author[uky]{C.~Crawford}
\author[utk,uky]{K.~Craycraft}
\author[utk]{N.~Fomin}
\author[inu,uva]{J.~Fry}
\author[umn]{M.~Gericke}
\author[ornl,inu]{R.C.~Gillis}
\author[utk,ornl]{K.~Grammer}
\author[utk,ornl]{G.L.~Greene}
\author[utc]{J.~Hamblen}
\author[utk]{C.~Hayes}
\author[ncs]{P.~Huffman}
\author[ornl]{C.~Jiang}
\author[utk]{S.~Kucuker}
\author[umn]{M.~McCrea}
\author[ornl]{P.E.~Mueller}
\author[ornl]{S.I.~Penttil{\"a}}
\author[inu]{W.M.~Snow}
\author[uky]{E.~Tang}
\author[inu,lanl]{Z.~Tang}
\author[utk,ornl]{X.~Tong}
\author[lanl]{W.S.~Wilburn}

\address[utk]{University of Tennessee, Knoxville, TN 37996, USA}
\address[mit]{Massachusetts Institute of Technology, Cambridge, MA 02139, USA}
\address[uva]{University of Virginia, Charlottesville, VA 22904, USA}
\address[ornl]{Oak Ridge National Laboratory, Oak Ridge, TN 37831, USA}
\address[asu]{Arizona State University, Tempe, AZ 85287}
\address[udam]{Instituto de F\'{i}sica, Universidad Nacional Autónoma\'{o}noma de M\'{e}xico, Apartado Postal 20-364, 01000, Mexico}
\address[umi]{University of Michigan, Ann Arbor, MI 48104, USA}
\address[uky]{University of Kentucky, Lexington, Kentucky 40506, USA}
\address[inu]{Indiana University, Bloomington, IN 47405, USA}
\address[umn]{University of Manitoba, Winnipeg, MB, Canada R3T 2N2}
\address[utc]{University of Tennessee, Chattanooga, TN 37403 USA}
\address[ncs]{North Carolina State University, Raleigh, NC 27695}
\address[lanl]{Los Alamos National Laboratory, Los Alamos, NM 87545, USA}

\date. 

\begin{abstract}
Accurately measuring the neutron beam polarization of a high flux, large area neutron beam is necessary for many neutron physics experiments.  The Fundamental Neutron Physics Beamline (FnPB) at the Spallation Neutron Source (SNS) is a pulsed neutron beam that was polarized with a supermirror polarizer for the NPDGamma experiment.  The polarized neutron beam had a flux of $\sim10^9$ neutrons per second per cm$^2$ and a cross sectional area of 10$\times$12~cm$^2$.  The polarization of this neutron beam and the efficiency of a RF neutron spin rotator installed downstream on this beam were measured by neutron transmission through a polarized \he\  neutron spin-filter.  The pulsed nature of the SNS enabled us to employ an absolute measurement technique for both quantities which does not depend on accurate knowledge of the phase space of the neutron beam or the \he\ polarization in the spin filter and is therefore of interest for any experiments on slow neutron beams from pulsed neutron sources which require knowledge of the absolute value of the neutron polarization.  The polarization and spin-reversal efficiency measured in this work were done for the NPDGamma experiment, which measures the parity violating \gr\ angular distribution asymmetry with respect to the neutron spin direction in the capture of polarized neutrons on protons. The experimental technique, results, systematic effects, and applications to neutron capture targets are discussed.
\end{abstract}


\end{frontmatter}


\section{Introduction}

Experiments in fundamental neutron physics often search for physical effects which are so small that one must use intense neutron beams with the largest cross sectional area allowed by the facility to see them in a reasonable running time.  Many of these experiments also require the use of polarized neutrons to search for spin-dependent effects.  The measurement of both the absolute neutron polarization and the efficiency of spin reversal for such large cross sectional area polarized slow neutron beams therefore must be conducted with sufficient accuracy to isolate the physical effects of interest.  The subclass of experiments which motivated our work are searches for the very small spin-dependent asymmetries in hadronic parity violation.  For most current parity violation experiments in this field the uncertainty is strongly limited by neutron counting statistics: in fact many of the effects predicted by theory have not yet been resolved experimentally.  It is therefore essential that the polarizer used in these experiments possess a high efficiency, can accommodate a large neutron beam cross sectional area, and operate continuously over long time periods of several months.  At the moment there are two practical choices for neutron polarizer technology for this work: supermirror neutron polarizers and neutron spin filters based on either polarized \he\ or polarized protons \cite{ZIMMER199962}. 

The differences of certain properties of these polarizers such as the radiation backgrounds, time stability, ability to reverse the beam polarization, and phase space uniformity of the beam polarization are obviously very important considerations for experiments.  An example of this is the NPDGamma experiment, which aims to measure the parity-violating \gr\ angular distribution asymmetry from polarized neutron capture on protons to an uncertainty of $1\times 10^{-8}$.  The \gr\ asymmetry is estimated to be $\sim 5\times 10^{-8}$ or less \cite{Desplanques1980,Wasem2011}, so the main challenge of the polarizer is to enable the high statistical precision since neutron polarization measurements better than 20\% are readily achievable.  A polarized \he\ spin filter was successfully used as a neutron polarizer for the initial phase of the NPDGamma experiment at LANSCE \cite{Rich2002,Chupp2007,Gericke2009,Gericke2011}.  The spatial uniformity of the polarized neutron beam phase space produced by such a spin filter, the ability to easily flip the polarization of the \he\ in the spin filter by NMR, and the absence of \grs\ from neutron capture in \he\ all contributed to this choice.  However, it was discovered that the high instantaneous neutron flux in the polarizer interfered with the long-term stability of the \he\ polarization produced in the spin-exchange optical pumping process, which occurred inside the neutron beam region \cite{Sharma2008}.  In addition, it was observed that potential sources of systematic errors in the experimental apparatus were very well suppressed by downstream spin reversal.  For the installation of the NPDGamma experiment at the FnPB at the SNS, whose much larger flux would have made this stability problem worse, a multichannel supermirror polarizer was used instead. This polarizer operates with a higher efficiency than present \he\ neutron spin filters and has a polarized neutron flux of $\sim 10^9$ neutrons per second per cm$^2$ \cite{Mason2006,Fomin2015}.  It is the measurement of the polarization from this device that is described in this paper.  

The neutron polarization was determined from transmission measurements through a polarized \he\ neutron spin filter.  A comprehensive review of the science behind optically pumped \he\ including its previous uses as a neutron spin filter can be found in \cite{gentile2016optically}.  This method takes advantage of the spin dependence in the capture cross section of neutrons on polarized \he\ and of the pulsed nature of the neutron beam at the SNS, which allows the neutron wavelength to be determined from time-of-flight analysis.  The pulsed neutron beam is also essential to our method of reversing the neutron polarization with high efficiency, and it allows the neutron polarization and spin-reversal efficiency to be determined independent of knowledge of the \he\ polarization \cite{Greene1995}.  To explain this, we review the physics of the operation of \he\ neutron spin filters below. 

\section{\he\ Polarimetry Principles}\label{sec:3he}

Neutrons readily capture on \he\ via the reaction $^{3}He + n \rightarrow p + {^{3}H}$.  The capture cross section is proportional to the neutron wavelength $\lambda$ such that the transmission of N$_0$ neutrons through an unpolarized sample of \he\ is 
\begin{linenomath*}
\begin{equation}
T_0=N_0e^{-n l\sigma_0 \frac{\lambda}{\lambda_0}},
\label{eq:unpol_trans}
\end{equation}
\end{linenomath*}
where $n$ is the atom density of \he, $l$ is the length of the \he\ spin filter, and $\sigma_0$ = 5316 b is the capture cross-section of $\lambda_0$ = 1.798 \AA\ neutrons \cite{ENDF2006}.  The capture cross section is spin dependent such that capture only occurs in the singlet state, i.e. neutron and \he\ spins are anti-aligned, and experimental measurements of capture into the triplet state are consistent with zero \cite{Huber2014}.  The transmission of spin up ($\uparrow$) and spin down ($\downarrow$) neutrons through a polarized \he\ spin filter with polarization $P_{He}$ is
\begin{linenomath*}
\begin{align}
T_\uparrow &=N_{\uparrow,0}e^{-nl\sigma_0 \frac{\lambda}{\lambda_0} (1-P_{He})} \label{eq:pol_up} \\
T_\downarrow &=N_{\downarrow,0}e^{-nl\sigma_0 \frac{\lambda}{\lambda_0} (1+P_{He})}.
\label{eq:pol_down}
\end{align}
\end{linenomath*}
For an unpolarized neutron beam, $N_{\uparrow,0}=N_{\downarrow,0}=\frac{1}{2}N_0$, and the total transmission of an unpolarized neutron beam through polarized \he\ is the sum over both neutron spin states
\begin{linenomath*}
\begin{align}
T = T_\uparrow+T_\downarrow = N_0e^{-nl\sigma_0 \frac{\lambda}{\lambda_0} }\cosh(nl\sigma_0 \frac{\lambda}{\lambda_0} P_{He}). \label{eq:pol_tot}
\end{align}
\end{linenomath*}

The transmission of a polarized neutron beam with a polarization of $P_n(\lambda)$ through polarized \he\ can be calculated for each neutron spin state from Eqs.~\ref{eq:pol_up}~\&~\ref{eq:pol_down} by making the neutron polarization implicit in the initial transmission coefficients, so that 
\begin{linenomath*}
\begin{align*}
T_\uparrow &= N_0 \frac{1+P_n}{2}e^{-nl\sigma_0 \frac{\lambda}{\lambda_0} (1-P_{He})}\\
T_\downarrow &= N_0 \frac{1-P_n}{2}e^{-nl\sigma_0 \frac{\lambda}{\lambda_0} (1+P_{He})}.
\end{align*}
\end{linenomath*}
The total transmission of a polarized neutron beam through polarized \he\ is the sum of the transmission of these two spin states:
\begin{linenomath*}
\begin{align}
T &= N_0e^{-nl\sigma_0 \frac{\lambda}{\lambda_0} }\cosh(nl\sigma_0 \frac{\lambda}{\lambda_0} P_{He}) \nonumber \\
&\quad \times \left[1+P_n \tanh(nl\sigma_0 \frac{\lambda}{\lambda_0} P_{He})\right].
\label{eq:trans}
\end{align}
\end{linenomath*}
If $P_{He}=0$, the transmission through unpolarized \he\ is found to be independent of the neutron polarization and Eq.~\ref{eq:trans} reduces to Eq.~\ref{eq:unpol_trans}. 

\subsection{Spin-Rotator Efficiency}\label{sbs:sf_eff}

On the FnPB the neutron spins are rotated by $\pi$ radians (spin-rotated) with an efficiency $0\le \epsilon_{sr}\le 1$ on a pulse-by-pulse basis.  We use the phrase ``spin-rotated''€ rather than ``spin-flipped'' to distinguish between two different modes of spin reversal: one in which the kinetic energy of the polarized neutron beam is unchanged (spin-rotated) and the other to denote the case when the kinetic energy of the neutron beam is changed (spin-flipped). It was important for us to employ the former type of spin reversal in NPDGamma to avoid potential sources of systematic error.  When the neutron spins are reversed, the magnitude of the neutron polarization becomes $(1-2\epsilon_{sr})P_n$.  The transmission through polarized \he\ with the spin-reversed neutrons can be determined similar to Eq.~\ref{eq:trans} by using the reversed polarization value $(1-2\epsilon_{sr})P_n$ such that
\begin{linenomath*}
\begin{align}
T_{sr} &= N_0e^{-nl\sigma_0 \frac{\lambda}{\lambda_0} }\cosh(nl\sigma_0 \frac{\lambda}{\lambda_0} P_{He}) \nonumber \\
&\quad \times \left[1+(1-2\epsilon_{sr})P_n \tanh(nl\sigma_0 \frac{\lambda}{\lambda_0} P_{He})\right].
\label{eq:withSF}
\end{align}
\end{linenomath*}
The spin-reversal efficiency can be calculated from four transmission measurements: two measured with the initial neutron spin state for both \he\ polarization states, $T$ and $T^{afp}$, and another two measured with the neutrons spin-reversed, $T_{sr}$ and $T^{afp}_{sr}$.  $T^{afp}$ and $T^{afp}_{sr}$ are transmission measurements where the \he\ polarization was reversed by adiabatic fast passage (AFP).  How AFP spin reversal is implemented and the AFP efficiency is accounted for are discussed in section~\ref{sec:setup}.  Two ratios are determined from these transmission measurements:
\begin{linenomath*}
\begin{align*}
\frac{T^{afp}-T}{T^{afp}+T} &= P_n \tanh\left(nl\sigma_0 \frac{\lambda}{\lambda_0} P_{He} \right) \\
\frac{T^{afp}_{sr}-T_{sr}}{T^{afp}_{sr}+T_{sr}} &= (1-2\epsilon_{sr})P_n \tanh\left(nl\sigma_0 \frac{\lambda}{\lambda_0} P_{He} \right).
\end{align*}
\end{linenomath*}
The ratios are equal to the neutron polarization after transmission through the supermirror polarizer and the polarized \he\ spin filter with the initial and spin-reversed neutron polarization respectively.  The spin-reversal efficiency can be calculated from these ratios as
\begin{linenomath*}
\begin{equation}
\epsilon_{sr} = \frac{1}{2}\left(1-\frac{\frac{T^{afp}_{sr}-T_{sr}}{T^{afp}_{sr}+T_{sr}}}{\frac{T^{afp}-T}{T^{afp}+T}}\right).
\label{eq:e_sf}
\end{equation}
\end{linenomath*}

\subsection{Neutron Polarization}\label{sbs:Npol}

The neutron polarization $P_n$ can be calculated from three transmission measurements: transmission through unpolarized \he\ (T$_0$ given by Eq.~\ref{eq:unpol_trans}) and transmission through polarized \he\ for both neutron polarization states (Eq.~\ref{eq:trans} and Eq.~\ref{eq:withSF}).  The ratio of the neutron transmission through polarized and unpolarized \he\ eliminates various forms of systematic error in the measurement that can come from nonideal properties of the transmission detector.  The ratios with the initial neutron spin state and reversed spin state are represented by R=T/T$_0$ and R$_{sr}$=T$_{sr}$/T$_0$ respectively, such that
\begin{linenomath*}
\begin{align*}
R &= \cosh\left(nl\sigma_0 \frac{\lambda}{\lambda_0} P_{He} \right)\left[1+P_n \tanh\left(nl\sigma_0 \frac{\lambda}{\lambda_0} P_{He} \right)\right] \\
R_{sr} &= \cosh\left(nl\sigma_0 \frac{\lambda}{\lambda_0} P_{He} \right) \\
&\quad \times \left[1+(1-2\epsilon_{sr})P_n \tanh\left(nl\sigma_0 \frac{\lambda}{\lambda_0} P_{He} \right)\right].
\end{align*}
\end{linenomath*}

By using the hyperbolic identity $1-\tanh^2x=$ sech$^2x$, the neutron polarization P$_n$ can be solved for as a function of the \he\ transmission ratios, R and R$_{sr}$, and the spin-reversal efficiency $\epsilon_{sr}$, such that
\begin{linenomath*}
\begin{equation}
P_n = \frac{R-R_{sr}}{\sqrt{\left[(2\epsilon_{sr}-1)R+R_{sr}\right]^2-4\epsilon_{sr}^2}}.
\label{eq:npol}
\end{equation}
\end{linenomath*}
An important implication of the result in Eq.~\ref{eq:npol} is that the neutron polarization can be determined without any knowledge of the \he\ polarization or the physical properties of the \he\ spin filter because R and R$_{sr}$ simply represent ratios of transmission measurements.  This statement remains true as long as the \he\ polarization is unchanged during transmission measurements of different neutron polarization states, which is easily accomplished due to the long T1 relaxation time of \he.

\section{NPDGamma Apparatus}

\subsection{Beamline}\label{sbs:beamline}

\begin{figure*}
         \centering
         \includegraphics[width=0.8\textwidth]{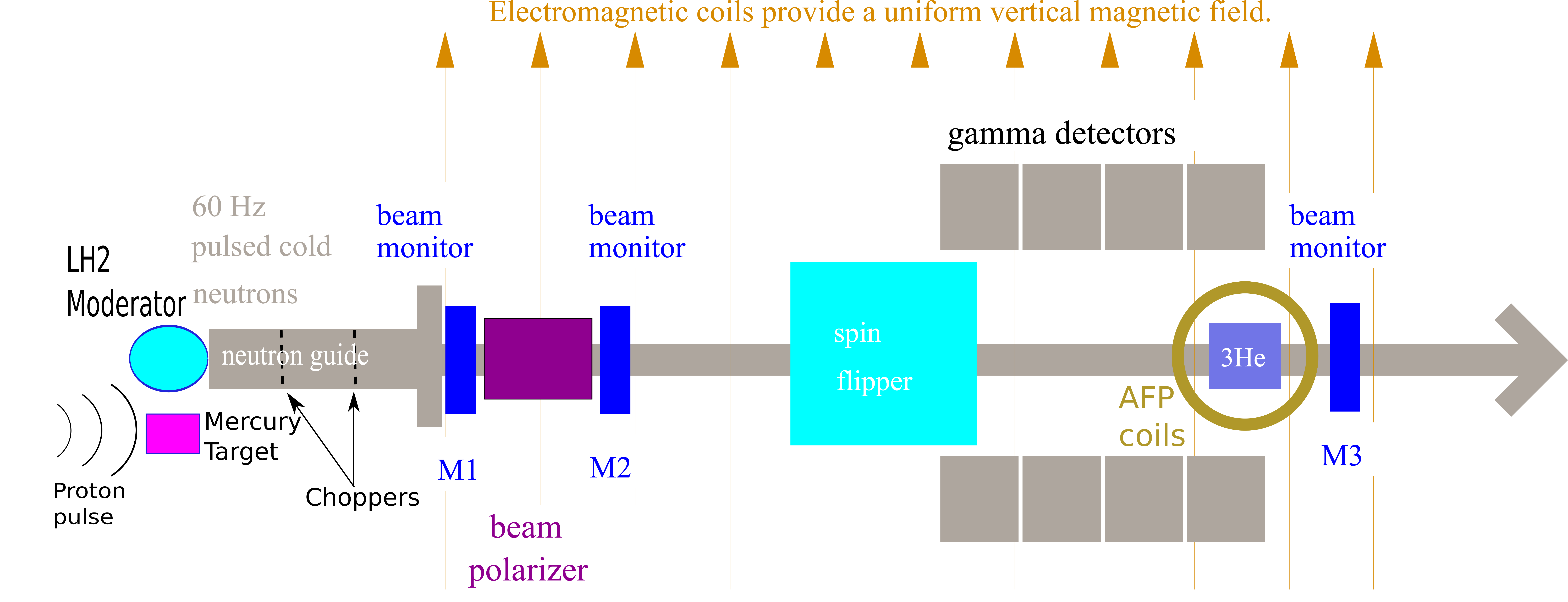}
         \caption[NPDGamma experiment]{The FnPB at the SNS with the NPDGamma Experiment configured for polarimetry measurements.}
         \label{fig:fnpb}
 \end{figure*}

The measurements described in this paper were taken at the FnPB with several components of the NPDGamma apparatus \cite{Fomin2015}.  A schematic of the beamline and the NPDGamma experiment are shown in Fig.~\ref{fig:fnpb}.  The pulsed neutron beam was produced by spallation from a proton beam on a Hg target at 60 Hz, and the neutrons were thermalized by a hydrogen moderator.  The neutron optical mirrors on the beamline are curved to eliminate direct line of sight to the moderator, and two neutron choppers are installed on the beamline to reduce the intensity of very slow neutrons from previous pulses from leaking into subsequent neutron time-of-flight windows (these are called frame overlap neutrons).  The beamline shutter transmits or blocks neutrons from the Hg spallation target to the NPDGamma experiment.

\subsection{Beam Monitors}\label{sbs:monitors}

Three neutron beam monitors are positioned along the experiment to monitor beam power and stability.  The beam monitors are \he\ proportional counters operated in current mode rather than as counting detectors because the high instantaneous neutron flux would create large deadtime and pileup errors for beam monitors of standard design and operation mode.  M3 is used to measure the transmission through the \he\ spin filter during polarization measurements, and M1 is used to normalize \he\ transmission measurements to the beam power.

\subsection{Supermirror polarizer}\label{sbs:smp}

The neutron beam is polarized by a multi-channel supermirror polarizer \cite{SwissN}.  The supermirror polarizer consists of 45 channels of uniformly spaced, curved supermirror panes with a radius of curvature of 14.8 m.  Each supermirror pane is a thin borosilicate glass substrate with a supermirror coating of alternating layers of Fe and Si evaporated onto both sides.  The materials in the supermirror polarizer had to be chosen to be compatible with the FnPB and NPDGamma experiment.  For example, the large neutron flux would have created unacceptable levels of Co-60 activation with a NiVCo/Ti supermirror polarizer, and a neutron absorbing Gd undercoating would have created a large \gr\ background that would dilute the NPDGamma \gr\ asymmetry measurement.  The supermirror polarizer operates in a 350 Gauss magnetic field that saturates the magnetization of the Fe layer, and the supermirror polarizer is placed inside a compensating magnetic field to reduce fringe fields \cite{Balascuta2012}.  The supermirror polarizer spin-filters one neutron spin-state that is able to reflect via Bragg diffraction on the Fe and Si layers.  However, neutrons with a small enough incident angle to reflect via total external refection are reflected for both spin states.  Because the angular divergences from each channel overlap, the neutron polarization varies horizontally across the beam cross section, and the experiment is positioned to optimized the incident neutron polarization.  The supermirror polarizer has a polarized neutron flux of $\sim 10^9$ neutrons per second per cm$^2$ at the exit and a cross-section of 12$\times$10 cm$^2$. 

\subsection{Magnetic Guide Field}\label{sbs:bfield}

The entire experiment downstream of the supermirror polarizer is inside of a uniform magnetic field of $\sim$9.4 Gauss to maintain the neutron polarization.  The walls, floor, and ceiling of the experimental cave are lined with low carbon steel to shield the experiment from extraneous magnetic fields and to improve the guide field uniformity by creating a flux return for the guide field.  The guide coil system also has shim coils to reduce field non-uniformities and allow small adjustments to the orientation of the guide field.  The magnetic field was mapped as described in \cite{Balascuta2012}, and temporal stability is continuously monitored by two fluxgate magnetometers.

\subsection{RF Spin Rotator}\label{sbs:rfsf}

The NPDGamma experiment uses a resonant RF spin rotator (RFSR), also known as a Rabi coil, to rotate the neutron spins $\pi$ radians, thereby isolating the small parity-violating signal and canceling several systematic uncertainties in the \gr\ asymmetry measurement \cite{Seo2008}.  The RFSR creates a magnetic field $\vec{B}_{RF}=B_{RF}\cos({\omega_{RF}}t)\hat{z}$ parallel to the neutron beam and on resonance such that the frequency of the RF magnetic field $\omega_{RF}$ is equal to the Larmor frequency of the neutrons in the guide field $\omega_0=\gamma_n B_0$.  The resonance frequency of the RFSR is set by hardware, where the impedance of the RFSR solenoid is the inductive component of a tuned resonance circuit, so the magnitude of the guide field is tuned to meet the resonance condition.

Within the RFSR neutron spins rotate about $\vec{B}_{RF}$ with a frequency of $\omega_1\propto B_{RF}$, and the probability of detecting the neutrons in a rotated spin state when the RFSR is on resonance is dependent on the time-of-flight through the RFSR t$_{sr}$ such that
\begin{linenomath*}
\begin{equation}
\epsilon_{sr} = \sin^2\left(\frac{\omega_1 t_{sr}}{2}\right).
\label{eq:idealSFeff}
\end{equation}
\end{linenomath*}
The time t$_{sr}$ is dependent on the neutron velocity, which for a pulsed neutron source is well defined by the time-of-flight from the moderator t$_{tof}$.  The spin-reversal efficiency is optimized for each neutron wavelength by varying the RF field amplitude as a function of t$_{tof}$ such that $B_{RF}\propto \frac{1}{t_{tof}}$ so that all neutrons that pass through the RFSR are rotated by $\pi$ radians.  The neutron spins are rotated with the sequence of spin reversal $\uparrow\downarrow\downarrow\uparrow\downarrow\uparrow\uparrow\downarrow$ to control linear and quadratic systematic effects.  The spin-reversal efficiency of the RFSR was measured with polarized \he\ as part of the polarimetry measurements.

\subsection{CsI(Tl) Detector Array}\label{sbs:targets}

The CsI(Tl) detector array measures the \grs\ emitted from polarized neutron capture on protons in a 16 K parahydrogen target \cite{Santra2010a,Grammer2015}.  The parahydrogen target is necessary because orthohydrogen has an incoherent scattering cross section about 50 times larger than the neutron capture cross section, and orthohydrogen contamination in the target would quickly depolarize and scatter the neutron beam.  However, the parahydrogen state is 14.7 meV lower energy than orthohydrogen, so for low energy neutrons and a 16 K target temperature, the total scattering cross section for neutrons is suppressed and the neutron polarization is preserved during scattering.  Additional \gr\ measurements with aluminum, chlorine, and boron targets are used to characterize the experiment.  The neutron flux and capture rates deduced from the \gr\ measurements are used to make appropriate polarization corrections to the parity violating \gr\ asymmetry from each target.  These corrections will be discussed later.

\section{Polarimetry Apparatus}\label{sec:setup}

\begin{figure}
	\centering
	\includegraphics[width=1.0\columnwidth]{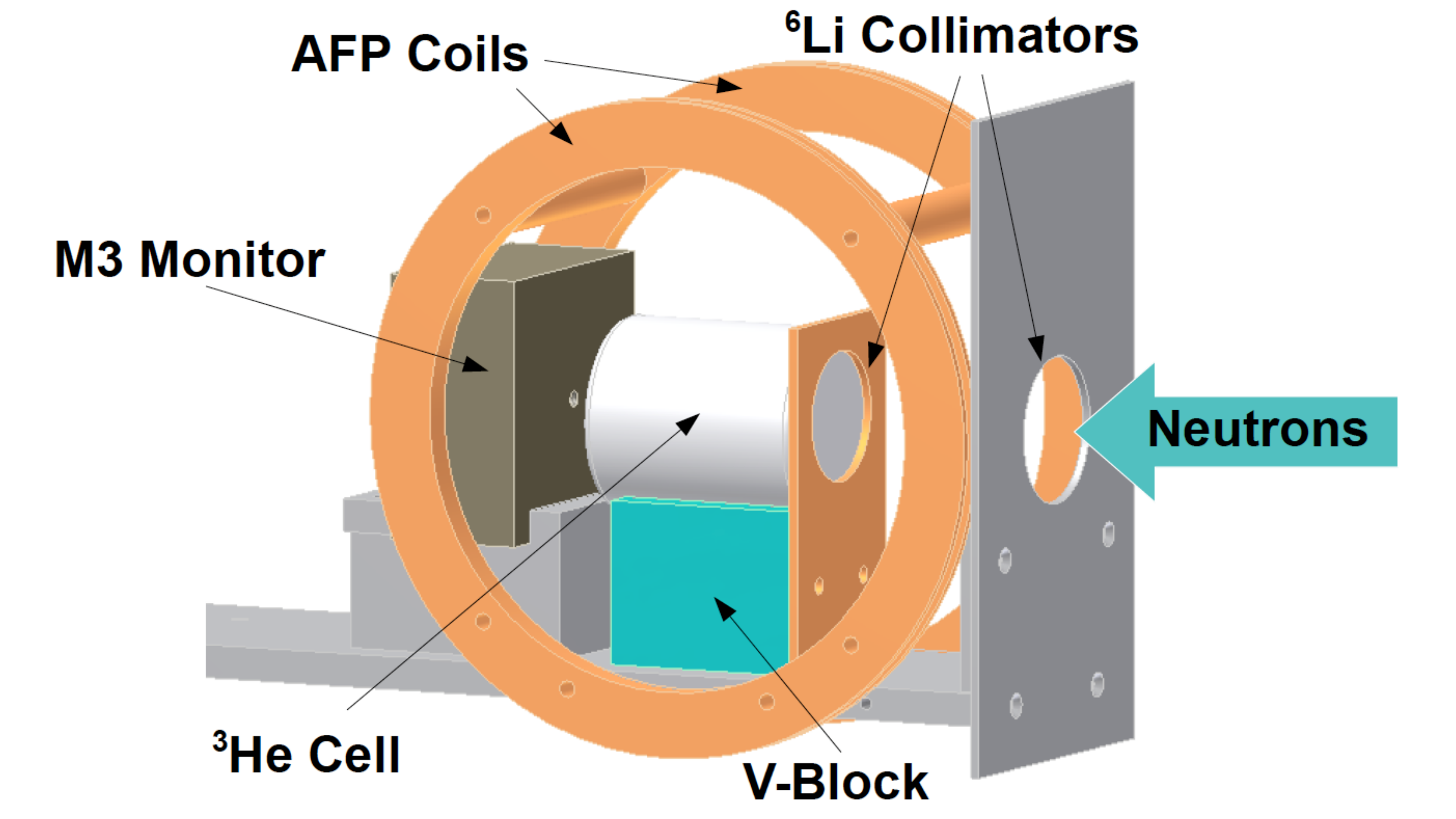}
	\caption[Polarimetry apparatus]{The polarimetry apparatus installed behind the NPDGamma experiment to measure the neutron polarization.}
	\label{fig:setup}
\end{figure}

A diagram of the polarimetry apparatus is shown in Fig.~\ref{fig:setup}.  The \he\ spin filter used for polarimetry is a hybrid K and Rb cell with a \he\ partial pressure of 1.31 $\pm$ 0.03 bars at 295 K as measured with neutron transmission.  The \he\ spin filter was fabricated and polarized with facilities available at ORNL \cite{JIANG2013191}.  The \he\ spin filter is cylindrical with an outside diameter of 7.5 cm and a length of 10.3 cm measured from the center of the circular faces.  The circular faces of the \he\ spin filter have some curvature as a result of the fabrication process.  This curvature combined with the nonuniform polarization profile of the beam could affect the polarization measurements, but conservative estimates show the effect is negligible.  The \he\ spin filter is polarized in a spin-exchange optical pumping station and transported to the neutron beamline in a transport coil, and \he\ polarizations between 60\% and 70\% have typically been available at the beamline.  Structural components of the polarimetry apparatus are nonmagnetic to maintain the \he\ polarization.

\begin{figure}[ht]
	\centering
	\includegraphics[width=1.0\columnwidth]{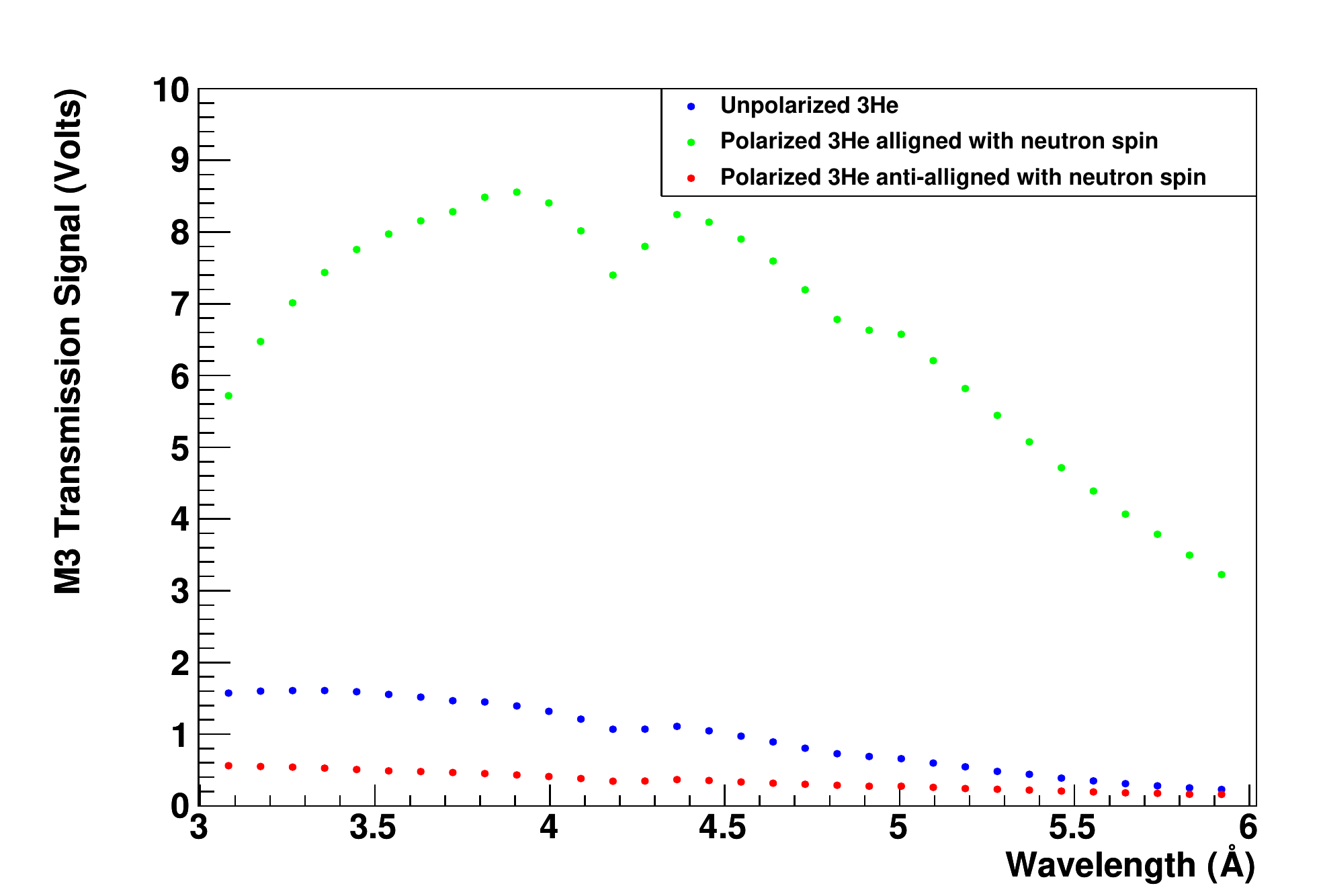}
	\caption[\he\ transmission signals]{Transmission signals through a polarized and unpolarized \he\ spin filter for both neutron spin states.  The edges in the plots at wavelengths 4.2 and 4.9 \AA\ are caused by Bragg scattering in aluminum.}
	\label{fig:spectrum}
\end{figure}

The polarized \he\ spin filter is placed in the beamline in front of the neutron monitor M3.  Neutron monitor signals are corrected for an electronic pedestal dependent on the gain of the monitor which is measured with the beamline shutter closed.  The transmission through polarized and unpolarized \he\ measured by M3 with the pedestal subtracted are shown in Fig.~\ref{fig:spectrum}.  For some configurations of the polarimetry apparatus, there is also a room neutron background in the M3 signals when the beamline shutter is open and details of the background determination are discussed later.

The \he\ spin filter is centered within a pair of Helmholtz coils that are used to reverse the \he\ polarization via AFP.  In this application of AFP, the guide field is held constant and the RF frequency in the AFP coils is swept across the \he\ nuclei's resonance frequency.  Measuring the transmission of both neutron spins states is necessary to determine the efficiency of the RFSR and to determine the neutron polarization independent of the RFSR.  The efficiency of an AFP reversal of the \he\ polarization was determined from measurements of the \he\ polarization via both neutron transmission and NMR, and the efficiency was found to be $\epsilon_{AFP}=0.972\pm 0.004$.  Whenever an AFP spin reversal was performed, three transmission measurements where taken with the \he\ polarization reversed in between each measurement, and the first and third transmission measurements were averaged.  The procedure sufficiently approximated the efficiency of one AFP-reversal $(1+0.972^2)/2=0.97239$ and reduced the uncertainty to neutron polarization and spin-reversal efficiency to less than 0.0001.  

A set of 5 cm diameter \li\ collimators defines a beam such that all transmitted neutrons pass through the \he\ spin filter.  In order to determine the beam-average neutron polarization and spin-reversal efficiency across the 12$\times$10 cm$^2$ beam cross section, the polarimetry apparatus was scanned horizontally and vertically in a 3-by-3 grid with a 4 cm separation between each grid point, and transmission measurements were taken at these nine positions.

After polarized \he\ transmission measurements, the \he\ spin filter is depolarized using the strong field gradients produced by a NdFeB permanent magnet.  Polarized neutron transmission measurements confirm that this method reliably and completely depolarizes the \he.

\section{Neutron Beamline and RFSR Models}

\begin{figure}[t]
	\centering
	\includegraphics[width=0.95\columnwidth]{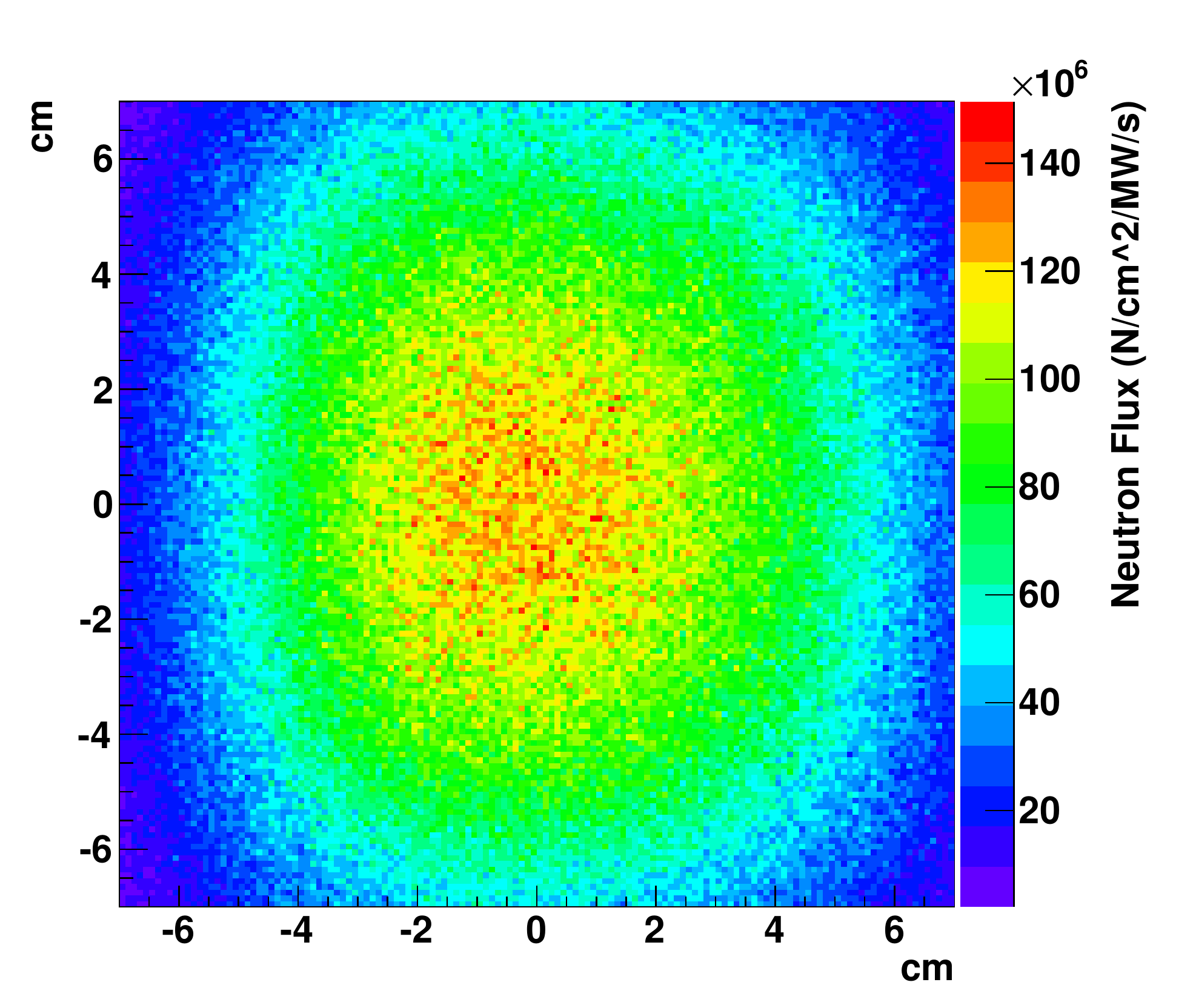}
	\caption[Simulation of the neutron flux]{The simulated neutron flux in a 14 cm by 14 cm cross section of the neutron beam 2.3 m downstream of the supermirror polarizer.  The neutron flux is normalized to the proton beam power in megawatts incident on the spallation target of the SNS.}
	\label{fig:model}
\end{figure}

\begin{figure}[h]
	\centering
	\includegraphics[width=1.0\columnwidth]{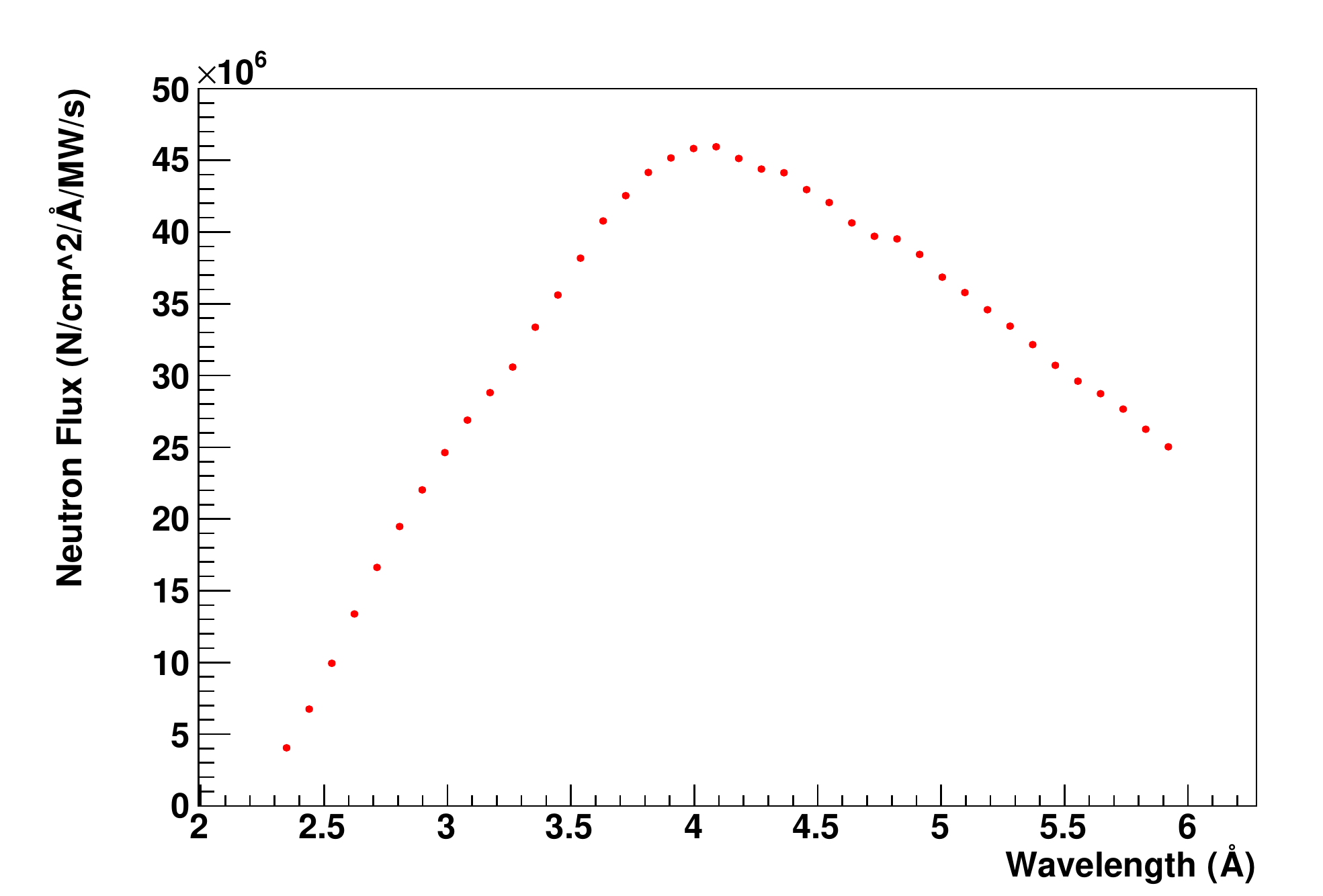}
	\caption[Simulated neutron flux versus wavelength]{The simulated neutron flux versus wavelength 2.3 m downstream of the supermirror polarizer.  The neutron flux is normalized to the wavelength bin size and the proton beam power.}
	\label{fig:model2}
\end{figure}

The FnPB from the moderator downstream to the supermirror polarizer is modeled with the neutron simulation package McStas \cite{McStas1999}.  After the supermirror polarizer, neutron positions are extrapolated by assuming that the neutrons travel in a straight line without any further interactions, which is a valid approximation because the beamline downstream of the SMP is lined with \li\ neutron shielding that nearly totally absorbs any scattered neutrons.  The simulated neutron flux through a cross section of the neutron beam 2.3 m downstream of the supermirror polarizer is shown in Fig.~\ref{fig:model}, and Fig.~\ref{fig:model2} displays the simulated neutron flux versus wavelength at the same beam position.  The beamline and supermirror polarizer models do not simulate Al Bragg scattering in the beamline. The effect of Al Bragg scattering appears in neutron monitor signals as two edges at 4.05~\AA\ and 4.68~\AA\ as seen in Fig.~\ref{fig:spectrum}.  

The RF field in the RFSR is described by modified Bessel functions, and the spin equations of motion of neutrons traversing the RF field are solved numerically \cite{Seo2008}.  The spin-reversal efficiency in the RFSR is position dependent, and therefore the average spin-reversal efficiency of 
\begin{figure}[h]
	\centering
	\includegraphics[width=0.95\columnwidth]{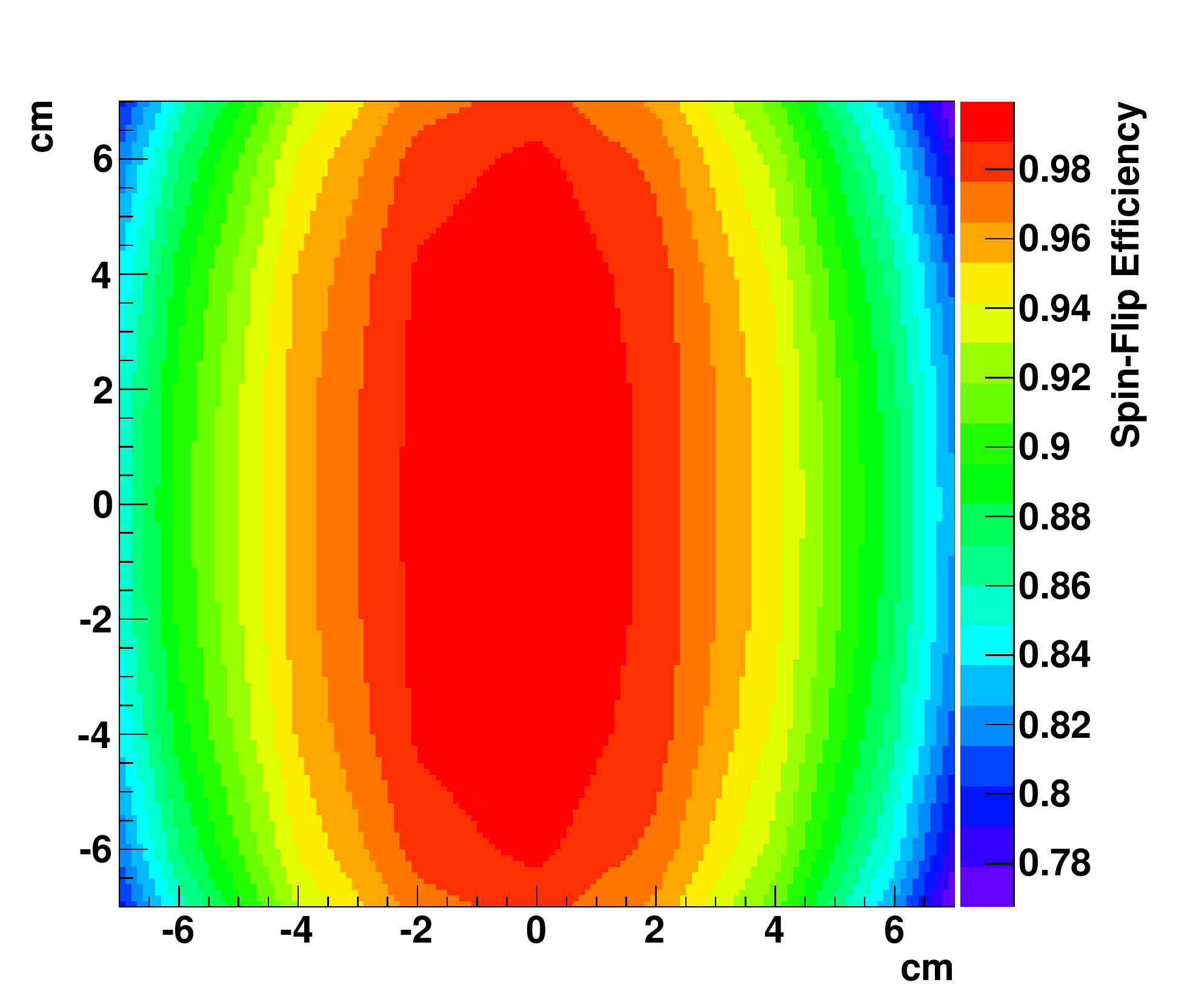}
	\caption[Simulation of the spin-reversal efficiency]{The simulated spin-reversal efficiency in a 14 cm by 14 cm cross section of the neutron beam 2.3 m downstream of the supermirror polarizer (1 m downstream of the RFSR).}
	\label{fig:SFmodel} 
\end{figure}
the neutron beam downstream of the RFSR is modeled by assigning each simulated neutron from the McStas model of the supermirror polarizer a spin-reversal efficiency based on its position in the XY-plane at the center of the RFSR.  The spin-reversal efficiency modeled across the beam cross section 1 m downstream of the RFSR is shown in Fig.~\ref{fig:SFmodel}.

\section{Polarimetry Results}\label{sec:results}

\subsection{Spin-Reversal Efficiency Measurements}\label{sbs:rfsf_eff}

\begin{figure}
        \centering
        \begin{subfigure}[b]{0.8\columnwidth}
        	\centering
        	\includegraphics[width=\columnwidth]{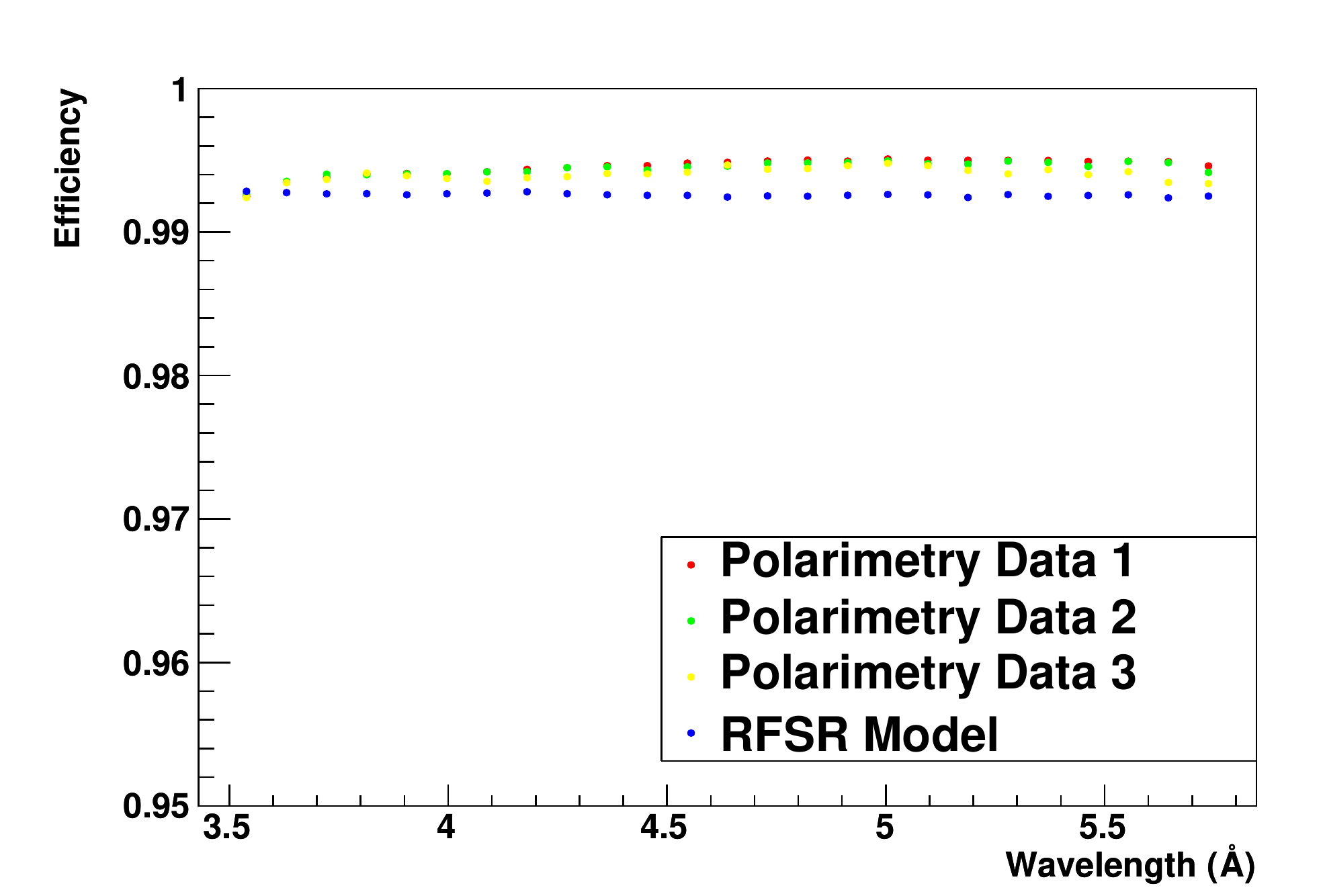}
        	\caption{}
        	\label{fig:sf_center}
        \end{subfigure}
        \begin{subfigure}[b]{0.8\columnwidth}			
        	\centering
        	\includegraphics[width=\columnwidth]{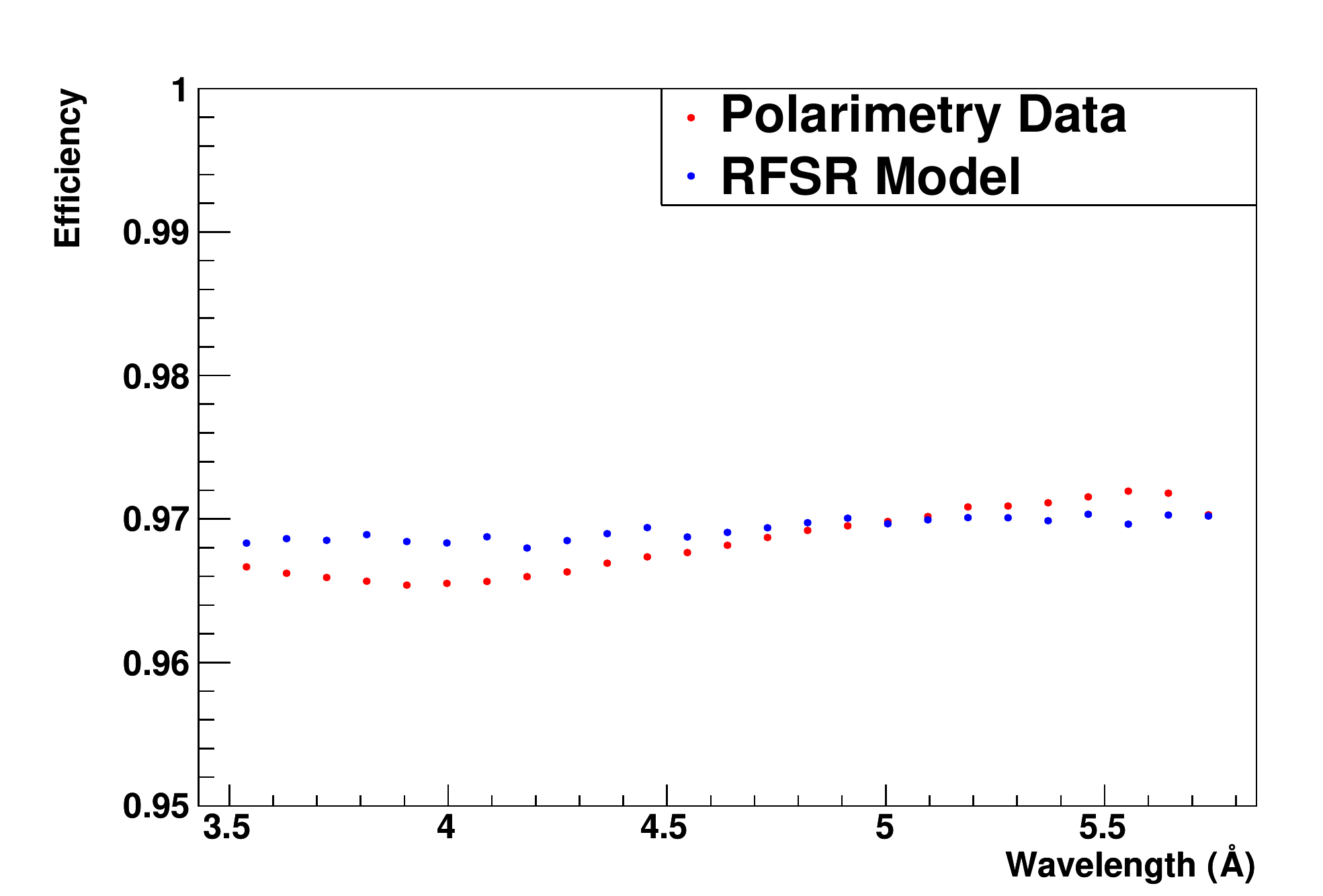}
        	\caption{}
        	\label{fig:sf_beamright}
        \end{subfigure}
        \caption[The spin-reversal efficiency]{The measured spin-reversal efficiency compared to the RFSR and beamline models at the center of the neutron beam (a) and 4 cm beam-right of center (b).}
        \label{fig:spin-flip}
\end{figure}

The spin-reversal efficiency was measured three times at the center of the neutron beam with three different \he\ polarizations and once 4 cm beam-right of center.  The measurements of the spin-reversal efficiency in these two positions are used to validate the model of the RFSR efficiency.  Transmission measurements were not taken in additional locations because the AFP \he\ spin reversals, which are necessary to determine the neutron spin-reversal efficiency, depolarized the \he\ and the calculated fields, which were verified by measurements, are trusted because the physics of the operation of the RFSR is well understood and depends only on the well-studied and understood interaction of the neutron magnetic moment with static and time-dependent magnetic fields.  The measured spin-reversal efficiencies are plotted in Figs.~\ref{fig:sf_center} \& \ref{fig:sf_beamright} along with the expected spin-reversal efficiencies determined from simulations with the beamline and RFSR models.  The average spin-reversal efficiency measured with a $\sim$5 cm diameter collimator is 0.995 $\pm$ 0.001 at the center of the neutron beam and 0.968 $\pm$ 0.002 beam-right of center in the wavelength range 3.5-5.8~\AA.  The measurements agree closely with model predictions to within NPDGamma's desired accuracy and validate the use of the model for extrapolating the measured spin-reversal efficiency to other neutron phase space regions.  The variation between measured spin-reversal efficiencies and the difference with model expectations are used to determine the uncertainty of the beam-average spin-reversal efficiency.

\subsection{Neutron Polarization Measurements}\label{sbs:polarization}

\begin{figure}
	\centering
	\includegraphics[width=1.0\columnwidth]{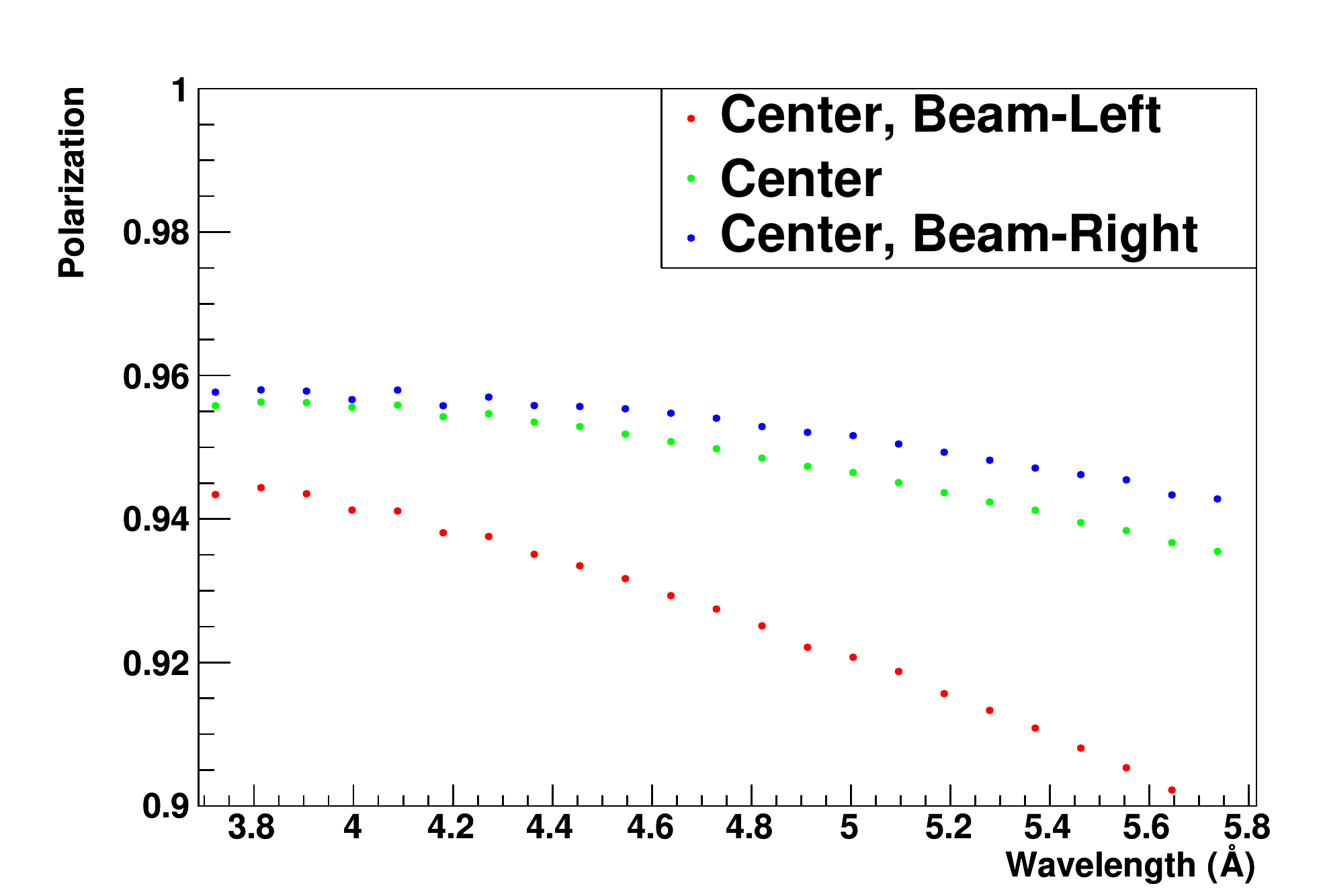}
	\caption[Neutron polarization at different beam positions]{The neutron polarization measured at different horizontal positions in the neutron beam.}
	\label{fig:beamcenter}
\end{figure}

The neutron polarization was measured at nine positions in a 3-by-3 grid in the neutron beam, so that the beam-average polarization can be determined from a weighted average of the nine polarization measurements.  Three polarization measurements at the same vertical position are plotted in Fig.~\ref{fig:beamcenter}.  The neutron polarization is lower on the beam-left side compared to the center and beam-right sections of the beam.  This phenomena is due to the superposition of the neutron flux from the 45 curved channels of the supermirror polarizer on each other 2.3 m downstream of the supermirror polarizer, where the \he\ spin-filter is positioned, so the polarization observed is a composite effect of the angular-dependence of the polarization from each channel.  The polarizations measured at the same horizontal position generally have the same wavelength dependence due to the approximate vertical symmetry of the supermirror polarizer.  The manufacturing process of the supermirror polarizer also creates some spatial nonuniformities that led to minor variations in performance across the supermirror polarizer.  These neutron polarization phase space nonuniformities do not affect the NPDGamma asymmetry measurement because of the use of the RFSR.

\begin{figure}
        \centering
        \includegraphics[width=1.0\columnwidth]{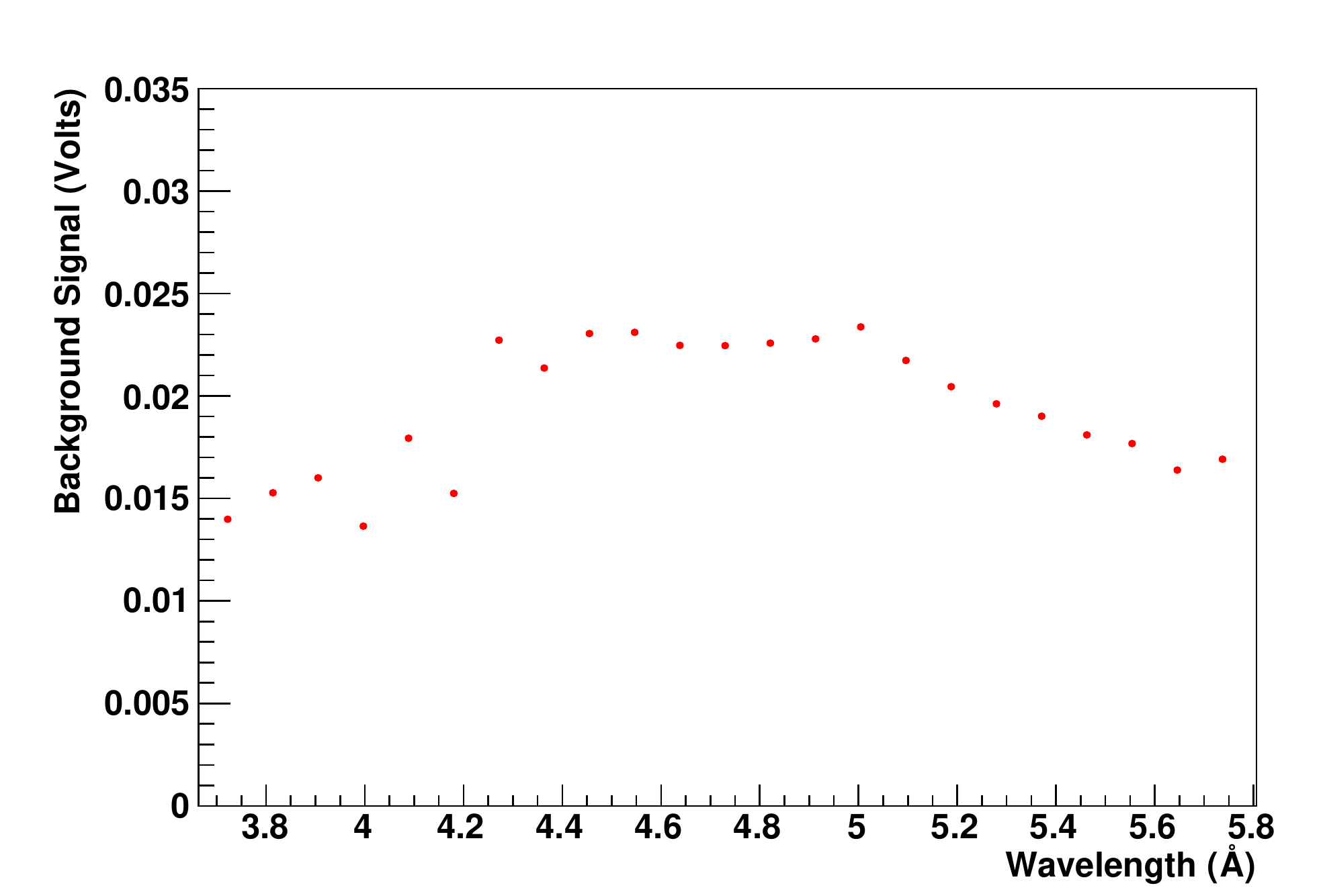}
        \caption[Neutron room background]{The neutron room background deduced from polarimetry measurements at different \he\ polarizations.}
        \label{fig:backg}
\end{figure}

The neutron polarization was measured in the center of the neutron beam with the \he\ spin filter at several polarizations to diagnose possible systematic effects correlated with the \he\ polarization.  When the M3 transmission signals were only corrected for the electronic pedestal, the measured neutron polarizations appeared to have a dependence on the \he\ polarization.  This phenomenon is consistent with a neutron room background that is only present when the beamline shutter is open.  The background in the M3 signals was not discovered until after the NPDGamma parahydrogen target was installed, so the size of the background is deduced by minimizing the variance in the neutron polarizations measured with different \he\ polarizations.  Since the spin-reversal efficiency is needed to determine the neutron polarization and the M3 signals need to be background-corrected in order to determine the spin-reversal efficiency, the background is calculated until the spin-reversal efficiency converged to a stable value, which required two iterations.  After correcting the M3 signals for the background in this way, the neutron polarization measurements with multiple \he\ polarizations were in agreement to within 1\%, and their variance is part of the uncertainty of the beam-average polarization calculation.  The background is plotted against wavelength in Fig.~\ref{fig:backg}.

Transmission measurements at off-center locations in the neutron beam were only taken with one \he\ polarization, so it is necessary to assume that the background does not vary with the position of the polarimetry apparatus.  This assumption is acceptable because the neutron polarization off-center is determined with transmission measurements through a relatively high \he\ polarization, which is less sensitive to the background. 

\begin{figure}
        \centering
        \includegraphics[width=1.0\columnwidth]{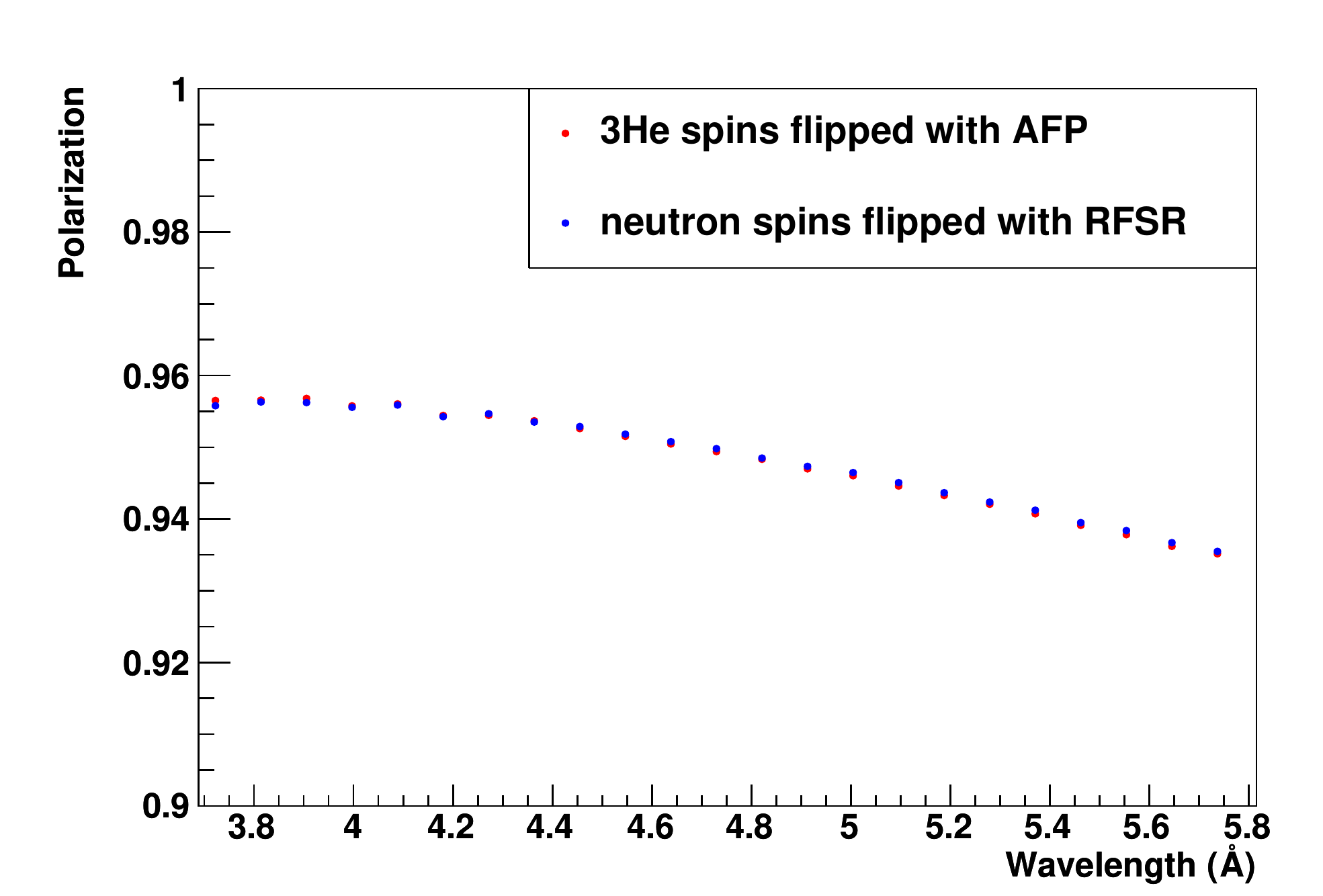}
        \caption[Neutron polarization determined with AFP]{The neutron polarization determined by either reversing the neutron spins with the RFSR or the \he\ spins with AFP.}
        \label{fig:afp_vs_rfsf}
\end{figure}

An equivalent method of determining the neutron polarization without flipping the neutron spins is to reverse the \he\ polarization and measure the transmission of one neutron spin state through both \he\ polarization states \cite{HEIL199756}.  The \he\ polarization is flipped by AFP.  By taking three transmission measurements with the \he\ polarization flipped in between each measurement and the first and third transmission measurements averaged, the AFP-flip efficiency of $\epsilon_{AFP}=0.972\pm 0.004$ leads to a correction to the neutron polarization of less than 0.01\%, which is negligible for our purposes.  The neutron polarizations determined by either flipping the neutron or \he\ spins are shown in Fig.~\ref{fig:afp_vs_rfsf}, and they are in agreement, which confirms both the measured neutron polarization and spin-reversal efficiency found using the RFSR.  

\subsection{Sources of Error}\label{sbs:errors}

\textit{Measurement Uncertainty}  The uncertainty of the neutron polarization is not limited by the neutron flux because the variance between measurements is much greater than the shot noise of counting statistics.  The uncertainty of the neutron polarization is calculated at each wavelength from the variation between measurements performed at several \he\ polarizations.  An upper limit of the standard deviation at all neutron wavelengths is 0.002, which includes the statistical uncertainty.  

\textit{Spin-Reversal Uncertainty}  The spin-reversal efficiency is 0.995 $\pm$ 0.001 in the center, and 0.968 $\pm$ 0.002 beam right of center.  The uncertainty of the spin-reversal efficiency at all locations off-center is assumed to be equal to the uncertainty beam-right of center.  The effect the uncertainty of the spin-reversal efficiency has on the uncertainty of the neutron polarization is calculated via propagation of errors and ultimately leads to a neutron polarization  uncertainty that is wavelength dependent.

\textit{\he\ Depolarization}  The T1 spin-lattice relaxation time of the \he\ spin filter in the guide field was measured with a series of free-induction decay (FID) NMR pulses that were taken over an 18 hour period at the FnPB.  The upper limit of the \he\ T1 relaxation time (assuming no depolarization due to the FID pulses) is 209.5 $\pm$ 5.4 hours.  The time between transmission measurements of different neutron spins states is sufficiently small that a 209 hour T1 relaxation time affects the neutron polarization measurement by less than 0.0001, which is completely negligible.

\begin{figure}
	\centering
	\includegraphics[width=1.0\columnwidth]{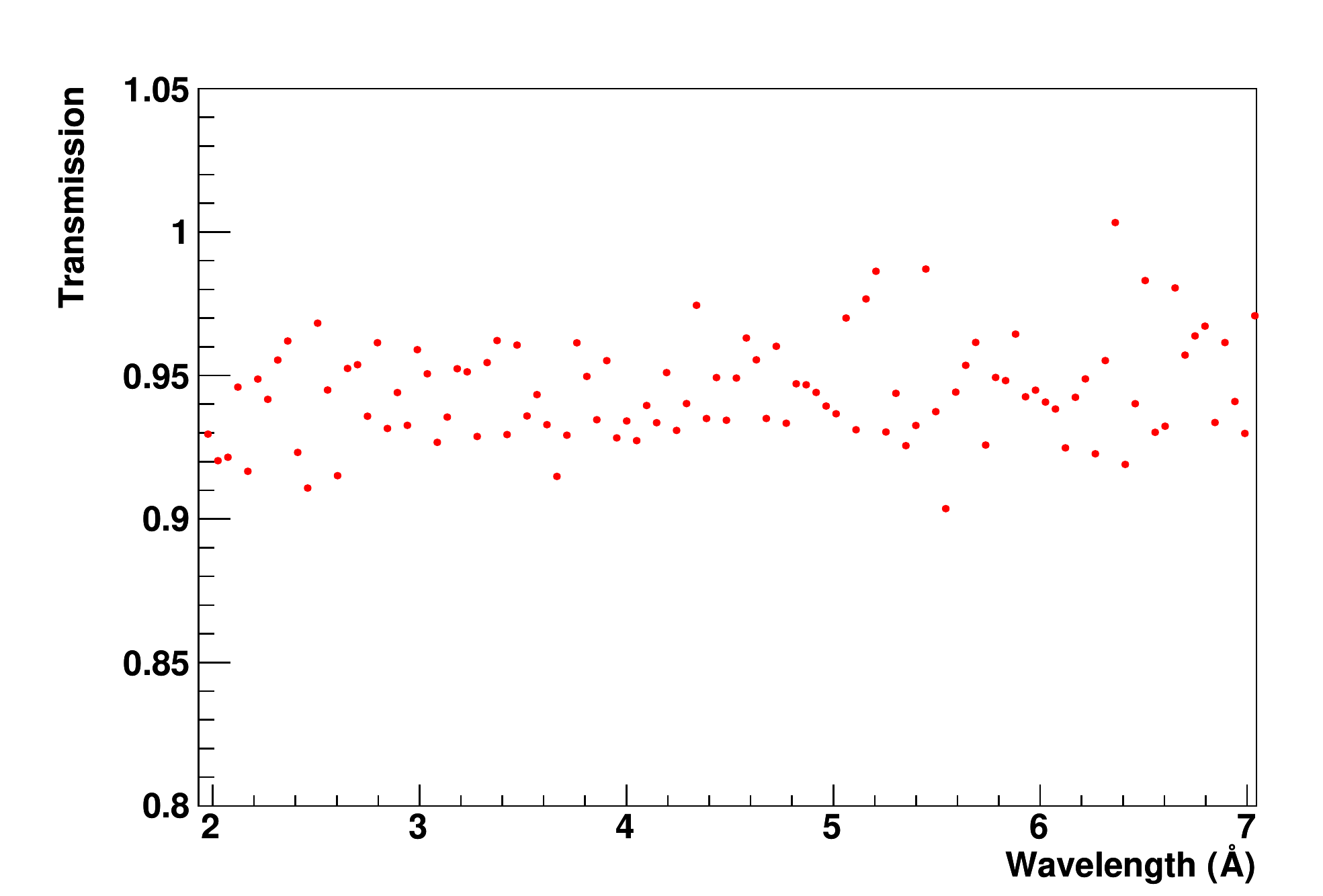}
	\caption[Neutron transmission through GE180 glass]{The neutron transmission through 3.175 mm of GE180 glass.}
	\label{fig:glass}
\end{figure}

\textit{Incoherent Scattering from GE180 Glass}  The transmission through an empty spin filter cell made of GE180 glass is approximated by the attenuation through an equivalent amount of GE180 glass.  Neutron transmission through 3.175 mm (1/8 inch) of GE180 glass was measured at beamline CG-1d at HFIR.  The attenuation through the glass is plotted in Fig.~\ref{fig:glass} and is relatively independent of wavelength, with a mean transmission of 0.94 $\pm$ 0.02.  This result is consistent with a previous measurement of neutron transmission through GE-180 glass \cite{Chupp2007}.  The walls of the \he\ spin filter used for polarimetry are about 1.5 mm thick, and the distance from the \he\ spin filter to the neutron monitor M3 is 13 cm.  Because of the small fraction of scattered neutrons, the distance to the M3 monitor, and the small incoherent scattering cross sections of the elements in GE180 glass, the effect of incoherent scattering in the glass walls of the \he\ spin filter on the neutron polarization is less than 0.0001.

\textit{Incoherent Scattering in Gas}  The \he\ spin filter contains 1.31 bar of \he\ and less than 0.1 bar of N$_2$.  In addition to being captured by \he, neutrons can scatter off \he\ or N$_2$, and incoherent scattering in the \he\ and N$_2$ gases can depolarize the neutrons just like incoherent scattering in the glass walls of the \he\ cell.  The cross sections for \he\ and N$_2$ are shown in Tab.~\ref{tab:NHe}.  The cross section for spin-flipping, incoherent scattering of neutrons on \he\ or N$_2$ is 0.01\% of the neutron capture cross section on \he, and since the ratio of detected to total neutrons that are scattered is 0.009, the depolarization of neutrons from scattering with \he\ and N$_2$ is negligible.  Neutron depolarization due to scattering in air and in the beamline windows is treated as an intrinsic property of the FnPB and is not considered here.

\begin{table}[t]
	\centering
	\caption[Neutron cross sections of \he\ and N$_2$]{Neutron cross sections (expressed in barns) of \he\ and N$_2$ measured with 0.0253 eV neutrons \cite{ENDF2006,mughabghab1984}.}
	\label{tab:NHe}
	\begin{tabular}{| l l l l l l |}
		\hline
		& $\sigma_\gamma$ & $\sigma_p$ & $\sigma_s$ & $\sigma_{coh}$ & $\sigma_{inc}$ \\
		\hline
		\he\ & 3.1e-5 & 5316 & 3.1 & 2.48 & 0.62 \\
		N & 0.07 & & 10.03 & 9.60 & 0.45 \\
		\hline
	\end{tabular}
\end{table}

\subsection{Beam-Average Values}\label{sbs:averageP}

The cross sectional area of the neutron beam at FnPB is exceptionally large in order to maximize neutron flux.  This creates the challenge of measuring the beamline's neutron polarization with a \he\ spin filter that does not intersect the entire beam.  Therefore, the neutron polarization and spin-reversal efficiency have to be measured for discrete areas of the neutron beam and then averaged.  The beam-average neutron polarization is the weighted average of the nine polarizations measured in a 3-by-3 grid in the beam and is calculated at each neutron wavelength using the expression
\begin{linenomath*}
	\begin{equation}
	\overline{P}_n(\lambda)=\frac{\sum\limits_{i} w_i(\lambda) P_i(\lambda)}{\sum\limits_{i} w_i(\lambda)}.
	\label{eq:wghtA}
	\end{equation}
\end{linenomath*}
The weights $w_i$ are proportional to the neutron flux at each position where the polarization was measured.  Wavelength dependent efficiency corrections can be ignored here because the average is being calculated at discrete wavelengths.  The transmission measured through unpolarized \he\ is used as the weight because it is proportional to the neutron flux at a given wavelength.

The beam-average spin-reversal efficiency is the weighted average of the spin-reversal efficiencies at the same nine positions as determined by
\begin{linenomath*}
	\begin{equation}
	\overline{\epsilon}_{sr}(\lambda)=\frac{\sum\limits_{i} w_i(\lambda) P_i(\lambda) \epsilon_i(\lambda)}{\sum\limits_{i} w_i(\lambda) P_i(\lambda)}.
	\label{eq:wghtE}
	\end{equation}
\end{linenomath*}
In addition to the relative neutron flux weight $w_i$, the spin-reversal efficiency at each position is weighted by the neutron polarization measured at each corresponding position.  The spin-reversal efficiency was measured in two positions in the beam where the polarization was measured, and it was determined in the other positions using the models of the supermirror polarizer and RFSR.  

In addition to the sources of measurement error discussed earlier, we must add the uncertainty due to the extrapolation of the measurements to the full beam.  The additional uncertainties of the beam-average polarization and spin-reversal efficiency are from an approximation of the full beam with nine sample measurements and the uncertainties of the neutron flux at the nine positions.  The nine sample regions do not intersect the entire beam cross section nor are they mutually exclusive.  The McStas model of the beamline is used to simulate the neutron flux through the nine circular sample regions and nine ideal, non-overlapping rectangular regions.  Two beam-average polarizations are calculated by weighting with the neutron flux through either the circular or rectangular sample regions, and the difference is taken to be the uncertainty associated with approximating the full beam with nine measurements. 

The uncertainty of the neutron flux at each sample position exists because the polarimetry apparatus had to be remounted between each polarized and unpolarized \he\ transmission measurement, which creates reproducibility errors.  The McStas model is used to simulate the neutron flux through a 5 cm diameter collimator at each measurement position, and based on the differences between simulated flux and the measured transmission, a conservative estimate of the uncertainty of the neutron flux is 0.025 (total flux normalized to 1).  The effect the uncertainty of the neutron flux has on the beam-average polarization is determined by varying the weights at each position by 0.025 and calculating the maximum change in the beam-average polarization, which is interpreted to be the uncertainty of the polarization due to the uncertainty of the weights. 

\begin{figure}
	\centering
	\includegraphics[width=1.0\columnwidth]{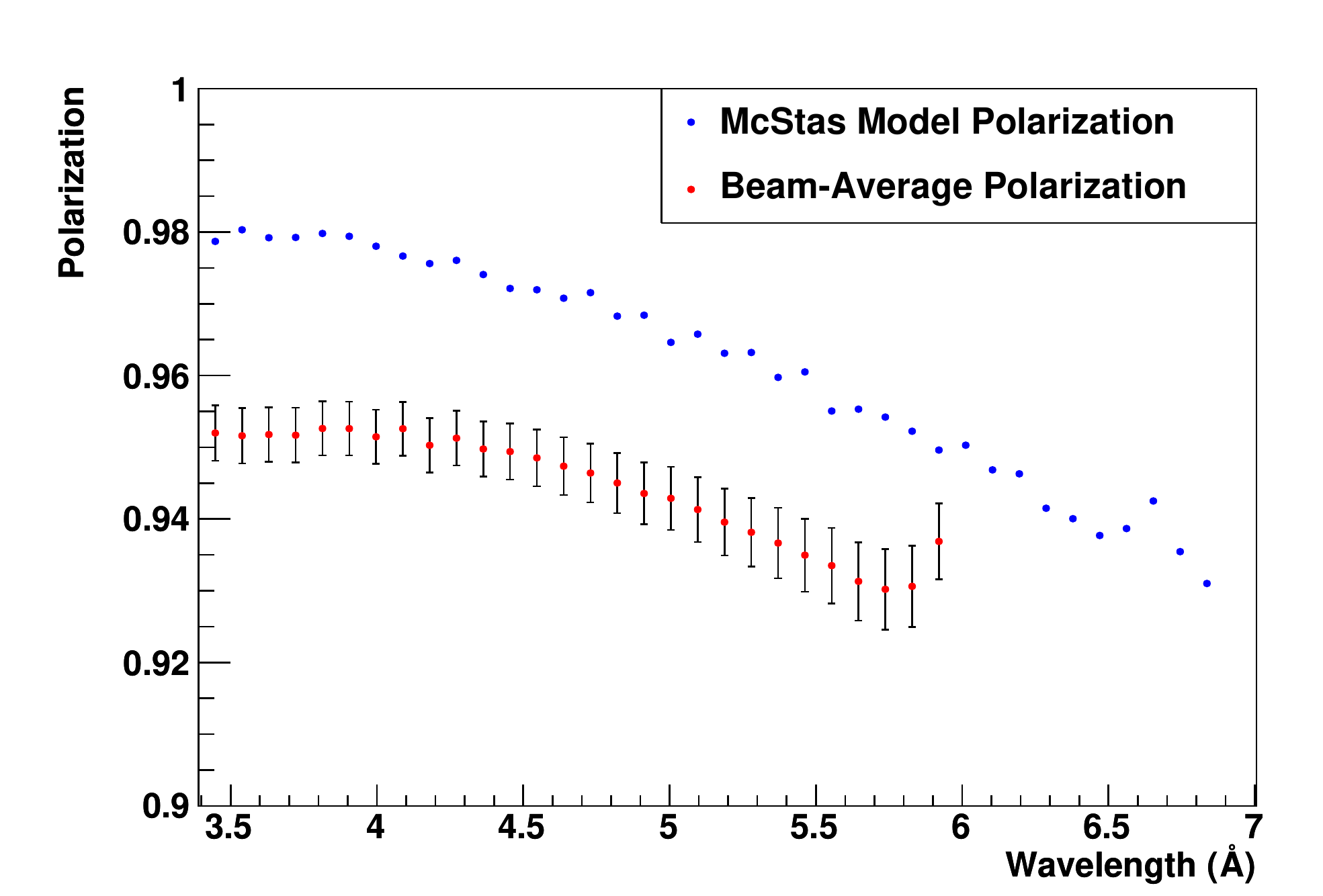}
	\caption[Beam-Average Polarization]{The measured beam-average polarization and the McStas model of the neutron polarization.}
	\label{fig:newpol_plot}
\end{figure}

The beam-average neutron polarization measured at the FnPB and the McStas model of the polarization of the FnPB are plotted in Fig.~\ref{fig:newpol_plot}.  Most sources of error are correlated; however, an increase in the uncertainty can be seen at larger wavelengths.  The absolute polarization value from the McStas model differs significantly from the measured polarization, but it does accurately predict the wavelength dependence of the neutron polarization.  This implies that the polarizing efficiency of the supermirror polarizer in the McStas model is too large but that McStas does accurately model neutron trajectories through the supermirror polarizer.  Possible reasons for the discrepancy between modeled and measured performance of the supermirror polarizer can include the surface reflectivity in the model (the model assumed NiVCo/Ti supermirror reflectivity curves), waviness of the 45 channels, and non-immediate neutron absorption in the supermirror polarizer.  The implication for error analysis that studies variation in the McStas polarization, such as the two sources previously discussed, is that the error will be slightly overestimated.

\section{Corrections for Neutron Capture Targets}\label{sec:capture}

\begin{table*}[t]
	\centering
	\begin{tabular}{| l l l |}
		\hline
		Source of Uncertainty & Neutron Polarization & Spin-Reversal Efficiency \\
		\hline
		Uncertainty of spin-rotator efficiency & 0.003 & N/A \\
		Magnetic field stability & N/A & 0.001 \\
		Variance of measurements/Neutron room background & 0.002 & 0.001 \\
		Approximating the beam with sample regions & 0.0001 & 0.002 \\
		Position-dependent flux for weighted average & 0.001 & 0.001 \\
		Wavelength-dependent flux for weighted average & 0.0005 & $<$0.0001 \\
		Uncertainty of target capture rate & 0.0006 & $<$0.0001 \\
		\he\ cell curvature & 0.0002 & 0.0002 \\
		AFP-flip efficiency & $<$0.0001 & $<$0.0001 \\
		Relaxation of \he\ polarization & $<$0.0001 & $<$0.0001 \\
		Incoherent scattering on GE180 glass & $<$0.0001 & $<$0.0001 \\
		Spin-flip scattering on \he\ and N$_2$ & $<$0.0001 & $<$0.0001 \\
		Neutron absorption on \he\ & $<$0.0001 & $<$0.0001 \\
		\hline
		Total uncertainty & 0.004 & 0.003 \\
		\hline
	\end{tabular}
	\caption[Polarization and spin-reversal efficiency uncertainties]{Uncertainties of the beam-average neutron polarization and spin-reversal efficiency from 3.7-5.8 \AA\ for the hydrogen target.}
	\label{tab:errors}
\end{table*}

The neutron polarization and spin-reversal efficiency are multiplicative corrections to the observed \gr\ asymmetry $A_{\gamma,obs}$ measured by the NPDGamma apparatus from neutron capture on parahydrogen, aluminum, and chlorine targets.  The physical asymmetry $A_{\gamma,phys}$ is calculated from the observed \gr\ asymmetry $A_{\gamma,obs}$ by
\begin{linenomath*}
	\begin{equation}
	A_{\gamma,phys} = \frac{A_{\gamma,obs}}{P_n \epsilon_{sr}},
	\label{eq:asymA}
	\end{equation}
\end{linenomath*}
where P$_n$ and $\epsilon_{sr}$ are the average neutron polarization and spin-reversal efficiency for the neutrons captured by the various targets as opposed to the values of the incident neutron beam.  This is necessary because the observed \gr\ asymmetry depends on the polarization of the neutrons captured.  

In order to correct the observed \gr\ asymmetry, the average polarization and spin-reversal efficiency of the neutrons captured is determined by a weighted average over the appropriate neutron wavelengths for each target with the expressions 
\begin{linenomath*}
\begin{equation}
\overline{P}_n=\frac{\sum\limits_{i} w(\lambda_i) \sigma(\lambda_i) P_n(\lambda_i)}{\sum\limits_{i} w(\lambda_i) \sigma(\lambda_i)}
\label{eq:avePn}
\end{equation}
\end{linenomath*}
and 
\begin{linenomath*}
\begin{equation}
\overline{\epsilon}_{sr}=\frac{\sum\limits_{i} w(\lambda_i) \sigma(\lambda_i) P_n(\lambda_i) \epsilon_{sr}(\lambda_i)}{\sum\limits_{i} w(\lambda_i) \sigma(\lambda_i) P_n(\lambda_i)},
\label{eq:aveSF}
\end{equation}
\end{linenomath*}
where $w(\lambda)$ is the neutron flux, $\sigma(\lambda)$ is the capture rate of the target, and P$_n$($\lambda$) and $\epsilon_{sr}$($\lambda$) are the beam-average polarization and spin-reversal efficiency of the neutron beam, which were measured from polarized \he\ transmission as described earlier.  

\begin{figure}
        \centering
        \includegraphics[width=1.0\columnwidth]{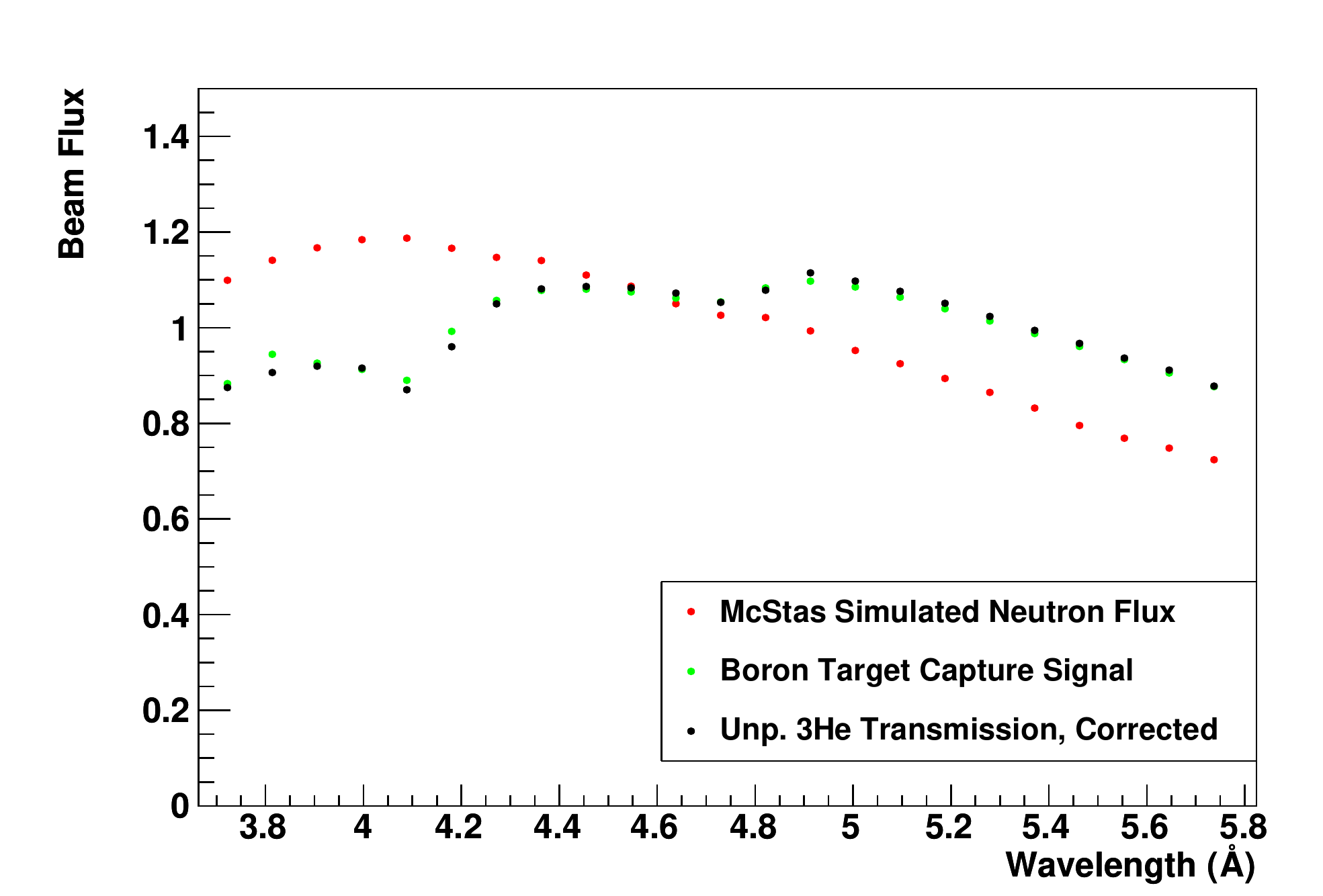}
        \caption[Neutron flux]{The beam-average neutron flux versus wavelength.  The neutron flux is normalized to an average value of 1.0 within the wavelength range of 3.7-5.8~\AA.}
        \label{fig:beamflux}
\end{figure}

The neutron polarization and spin-reversal efficiency are weighted by the neutron flux at each wavelength where the \he\ transmission was measured.  Weights proportional to the neutron flux were obtained from the McStas model of the supermirror polarizer, the \gr\ signal from neutron capture on a totally absorbing boron target, and the transmission measurements through unpolarized \he\ at the nine sample regions.  In order to determine the neutron flux from the transmission through unpolarized \he, the nine transmission measurements are summed and corrected for the neutron attenuation in the \he\ spin filter and the wavelength-dependent efficiency of the neutron monitor M3.  The neutron attenuation in the \he\ spin filter is given by Eq.~\ref{eq:unpol_trans}, and the efficiency of M3 is proportional to $e^{-\tau\lambda}(1-e^{-2\tau\lambda})$, where $\tau$ = 0.080 \AA$^{-1}$.  Wavelength-independent scaling corrections for each method of obtaining the neutron flux cancel during the averaging and are therefore ignored.  The results of each method of determining the neutron flux are normalized to 1.0 and plotted against wavelength in Fig.~\ref{fig:beamflux}.

The measured neutron flux determined from boron capture and transmission through unpolarized \he\ are consistent.  However, the simulated neutron flux differs because the McStas model of the supermirror polarizer does not include aluminum Bragg scattering in the beamline.  The neutron flux determined from transmission through unpolarized \he\ is used in the weighted average, and the variations in the average polarization and spin-reversal efficiency when weighted by the three alternative methods is incorporated into the uncertainty.

The capture rate is dependent on the target's neutron capture cross section and geometry as well as the properties of the surrounding experiment (e.g. CsI(Tl) detector array and \li\ shielding).  Therefore, the wavelength-dependent capture rate for each target is simulated in a MCNPX model of the NPDGamma experiment by counting the captures in the target per simulated source neutron over a range of neutron wavelengths.  The capture rates are also deduced from the \gr\ signals measured in the CsI(Tl) detector array for each target by comparing them to the measured neutron capture signal with a totally absorbing boron target.  Thus, the ratio of the \gr\ signal for each capture target to the boron capture signal is proportional to the capture rate as a function of wavelength.  Wavelength-independent corrections to the CsI(Tl) detector signals (e.g. the branching ratio of the neutron-capture reaction that produces a \gr, the energy of the emitted \grs, and the detector efficiencies) cancel during the averaging and are ignored.  The simulated and measured capture rates are in good agreement for all the targets, and the variations in the average polarization and spin-reversal efficiency when weighted by the two alternative methods is also incorporated into the uncertainty.

Neutron capture measurements took place over the course of days for chlorine, months for aluminum, and two years for parahydrogen, so additional polarimetry measurements were performed before each capture target was installed or at a maximum of once every two months to insure that the neutron polarization and spin-reversal efficiency did not vary with time.  Polarimetry results over the NPDGamma runtime showed that the experimental apparatus operated as expected with no detectable variations in its performance.  The two fluxgate magnetometers occasionally measured sudden fluctuations in the magnetic field on the order of $\sim$0.01 Gauss.  This adds a magnetic field stability uncertainty to the spin-reversal efficiency of 0.001 for the parahydrogen target only. 

The sources of uncertainty for neutron capture on the parahydrogen target are shown in Tab.~\ref{tab:errors}.  The sources of uncertainty for the aluminum and chlorine targets differ only slightly.  The polarization and spin-reversal efficiency used to correct the observed \gr\ asymmetry from neutron capture on the parahydrogen, aluminum, and chlorine targets are calculated with Eqs.~\ref{eq:avePn} \& \ref{eq:aveSF} using the weights discussed above and are displayed in Tab.~\ref{tab:results}.  

\begin{table}[t]
	\centering
	\begin{tabular}{| l l l |}
		\hline
		Target & Polarization & RFSR Efficiency \\
		\hline
		Parahydrogen & 0.943 $\pm$ 0.004 & 0.975 $\pm$ 0.003 \\
		Aluminum & 0.945 $\pm$ 0.004 & 0.979 $\pm$ 0.002 \\
		Chlorine & 0.946 $\pm$ 0.004 & 0.979 $\pm$ 0.002 \\
		\hline
	\end{tabular}
	\caption[Neutron polarization and spin-reversal efficiency for each target]{The average neutron polarization and spin-reversal efficiency from 3.7-5.8 \AA.  The aluminum and chlorine targets have higher polarization and spin-reversal efficiencies because the neutron beamline was more tightly collimated for these measurements.}
	\label{tab:results}
\end{table}

\section{Conclusion}\label{sec:conclusion}

The goal of the NPDGamma experiment is to make a high precision measurement of the parity violating \gr\ asymmetry in the capture of polarized neutrons on protons from the hadronic weak interaction.  In order to minimize systematic uncertainties so that the experiment is counting statistics limited, the neutron polarization and spin-reversal efficiency were successfully measured across the large cross section of the FnPB with neutron transmission measurements through a polarized \he\ neutron spin filter.  The uncertainty in the neutron polarization presented in this paper makes a negligible contribution to the uncertainty in the measured NPDGamma physics asymmetry.  This technique is likely to be useful in future experiments with pulsed polarized neutron beams such as in other hadronic parity violation searches, neutron scattering experiments, neutron beta decay experiments, and the SNS nEDM experiment.  In future applications, the neutron polarization measurement could be improved by reducing the neutron room background or by measuring the polarization with more precise sampling of the beam cross section.

\section*{Acknowledgements}

We would like to thank the ORNL glass shop and the University of Tennessee machine shop for their valuable contributions.  We also thank the management and staff of the Spallation Neutron Source for providing their support and for keeping the neutron source reliably running.  We gratefully acknowledge the support of the US Department of Energy Office of Nuclear Physics (Grants No. DE-FG02-03ER41258, DE-SC0008107, and DE-SC0014622), the US National Science Foundation (Grants No. PHY-1306942, PHY-1614545, NSF-0855610, and NSF-1205833), the Indiana University Center for Spacetime Symmetries, the Natural Sciences and Engineering Council of Canada (Grant No. SAPPJ 341289-2013), and the Canada Foundation for Innovation.

\section*{References}
\bibliography{citations}

\end{document}